# A NEW ALGORITHM FOR CONTACT ANGLE ESTIMATION IN MOLECULAR DYNAMICS SIMULATIONS

**Sumith YD**
Department of Mechanical and Aerospace Engg.,
Syracuse University, Syracuse, NY, USA

**Shalabh C. Maroo***
Department of Mechanical and Aerospace Engg.,
Syracuse University, Syracuse, NY, USA
Email: scmaroo@syr.edu

**ABSTRACT**

It is important to study contact angle of a liquid on a solid surface to understand its wetting properties, capillarity and surface interaction energy. While performing transient molecular dynamics (MD) simulations it requires calculating the time evolution of contact angle. This is a tedious effort to do manually or with image processing algorithms. In this work we propose a new algorithm to estimate contact angle from MD simulations directly and in a computationally efficient way. This algorithm segregates the droplet molecules from the vapor molecules using Mahalanobis distance (MND) technique. Then the density is smeared onto a 2D grid using 4$^{th}$ order B-spline interpolation function. The vapor liquid interface data is estimated from the grid using density filtering. With the interface data a circle is fitted using Landau method. The equation of this circle is solved for obtaining the contact angle. This procedure is repeated by rotating the droplet about the vertical axis. We have applied this algorithm to a number of studies (different potentials and thermostat methods) which involves the MD simulation of water.

## INTRODUCTION

It is quite common in the literature to use Berendsen[1] or Nose-Hoover[2] thermostat for liquid molecular dynamic simulations to study different thermodynamic properties of water. In this paper alongside with the new algorithm, we have investigated the effect of such thermostats on the contact angle of water with respect to wall heating algorithm [3].

In the past Bo Shi et al [4] have simulated the contact angle of water on top of Platinum surface with FCC 111. Their work was with simulating columbic potential with P3M method [5] and keeping the temperature constant using Berendsen thermostat [1]. Maruyama and et al [6] have also done contact angle studies of water on platinum with truncated potential.

Both researchers have used ZP potential for water platinum interaction. There are previous works on contact angle estimation from MD simulations. Erik et al [7] have come up with an idea of smearing the molecules into grid with Nearest Grid Point (NGP) scheme which we will discuss the drawback later in this paper. In another work Malani et al [8] have shown a new method to study contact angle from MD. Sergi et al [9] have calculated contact angle of water from MD using local averaging and fitting methods. Also there exists a wide range of studies done on contact angle estimation using MD simulations [10-13]. Most of these researchers have focused on finding contact angle and surface energy through different ways. However almost everyone uses NGP scheme which makes the prediction of liquid-vapor interface skeptical. Again a discussion about transient monitoring of contact angle is not found in the literature.

With the enormous amount of result data estimation of contact angle using image recognition algorithms or manual methods becomes almost impossible. There are not many algorithms which suit well for the direct post processing of molecular simulations. The paper is divided into two sections, first explaining the algorithm used for estimating contact angle; second the details of MD Simulation and models used.

## ALGORITHM DETAILS

Often MD simulation results of water and other liquids will have the information of co-existence of vapor and liquid. Segregating the liquid droplet or liquid from the vapor will avoid false detection of Solid-Liquid boundary (LVB) when using density filtering. Though this step is optional, it would give more appropriate data for estimating contact angle. For convenience the whole process is divided into 5 steps.

### Step 1: Sample preparation

The unwanted vapor molecules (outliers) in the data can be removed effectively using Mahalanobis Distance [17] technique.

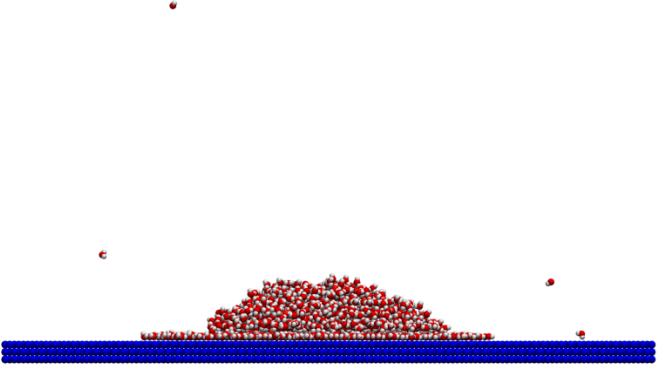

**Fig. 1. Equilibrated water droplet on platinum**

If $X_c$ is the $n \times 2$ column centered vector consisting of $(x - \bar{x}, y - \bar{y})$ data of n points then variance-covariance matrix $C_x$ is defined as

$$C_x = \frac{1}{(n-1)}(X_c)^T(X_c) \quad (1)$$

Then the Mahalanobis Distance (MND) is calculated as

$$MND_i = \sqrt{X_i C_x^{-1} X_i^T} \quad (2)$$

Where $X_i$ is the mean centered data of i$^{th}$ data point. From this list of MND we can neglect those data points with considerably high MND values.

The effectiveness of MND technique can be visualized from Fig. 2 and Fig. 3. The vapor molecules which would have been a hindrance to estimate the contact angle accurately was removed easily with MND method.

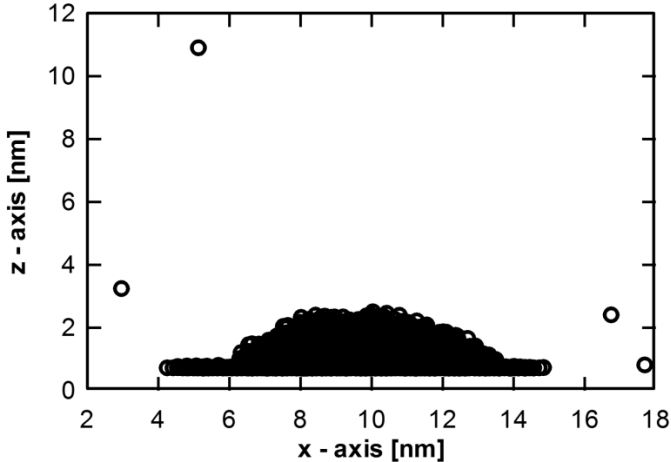

**Fig. 2. Center of mass data with vapor molecules**

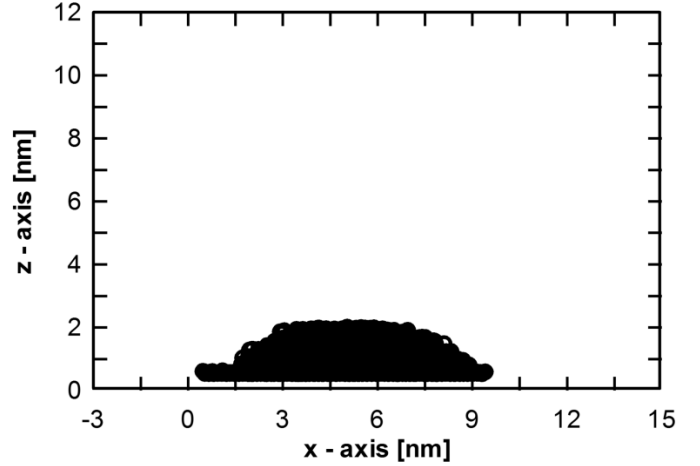

**Fig. 3. Data after removing the vapor molecules using MND**

### Step 2: Density grid using B-Splines

In this step a 2D mesh in XZ plane (see figure 2) is generated and the entire droplet data is projected and smeared into the grids using 4$^{th}$ order B-spline function. This makes it easier to find a smooth transition between the liquid core and the empty space.

The Nearest Grid Point (NGP) method is a traditional first order method normally researchers use to assign data to the grid points. Inspired from Hockney and Eastwood [18] the density of the molecules at every grid point using NGP scheme can be calculated by the below equation.

$$\rho_{ij} = \sum_{p=1}^{N} W\left(\frac{|x_i - x_p|}{dx}\right) * W\left(\frac{|y_i - y_p|}{dy}\right) \quad (3)$$

Where the weight function is defined as

$$W(x) = \begin{cases} 1, & x < \frac{1}{2} \\ 0, & otherwise \end{cases} \quad (4)$$

The B-spline method is inspired from Smooth Particle Mesh Ewald [19]. Reproducing the definition gives, for any real number $u$, let $M_2(u)$ denote the linear hat function given by $M_2(u) = 1 - |u - 1|$ for $0 \leq u \leq 2$ and $M_2(u) = 0$ for $u < 0$ or $u > 2$. For $n$ greater than 2, define $M_n(u)$ by the recursion

$$M_n(u) = \frac{u}{n-1} M_{n-1}(u) + \frac{n-u}{n-1} M_{n-1}(u-1) \quad (5)$$

For our case $n = 4$ and $u$ is in fractional coordinates. From Fig. 4 we can see that the B-spline scheme gives an upper hand in predicting the LVB.

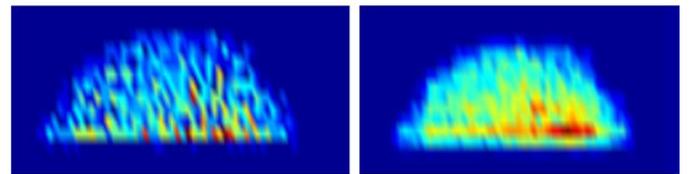

**Fig. 4. a) NGP method, b) B-Spline method**

### Step 3: Density based filtering

In this step the exact LVB is calculated based on threshold density bands. Mathematically this can be explained using Equations 6 and 7. This will remove the grid points with high densities which resemble the liquid region and low densities that resemble the vapor (noise).

$$\text{Threshold}_{MAX} = w1 * \text{Mean} \tag{6}$$

$$\text{Threshold}_{MIN} = w2 * \text{Mean} \tag{7}$$

$w1$ and $w2$ are weights which can be fine-tuned according to the data. For the present studies we have taken them as 2 and 0.5 respectively. Now, all the data which meets the below criteria will be used for further analysis.

$$\text{Threshold}_{MIN} \leq \text{Grid.value} \leq \text{Threshold}_{MAX} \tag{8}$$

This will lead to detection of the points as shown in Fig. 5.

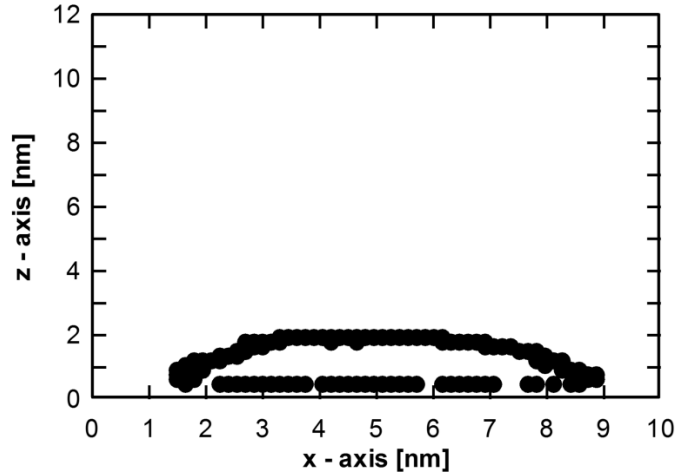

**Fig. 5. Density based filtered data**

### Step 4: Removing the wall effects

As one may notice, an unwanted lower set of contour is also detected due to the interface between water and platinum. This has to be taken care as explained next. Find the centroid of the data from the Step 3 and set the z coordinate as the top of the platinum plate. Then we radially move outwards from that point until we reach the boundary of data. This procedure is repeated for $0 \leq \theta \leq \pi$. This will help us to identify the most appropriate data. At the end we will be having a collection of points which forms the exterior of the droplet.

Fig. 6 shows this procedure graphically and Fig. 7 shows the detected droplet boundary. This will also eliminate the unwanted data points inside the droplet due to the occasional void formation due to the density fluctuations in the MD simulations. Also using a z-location based cutoff we can eliminate the unwanted monolayer or interface formed next to the platinum wall.

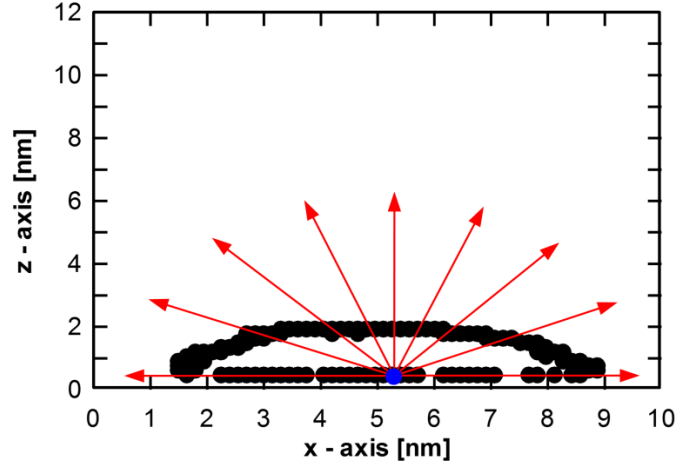

**Fig. 6. Radial rays from centroid to find the extremities**

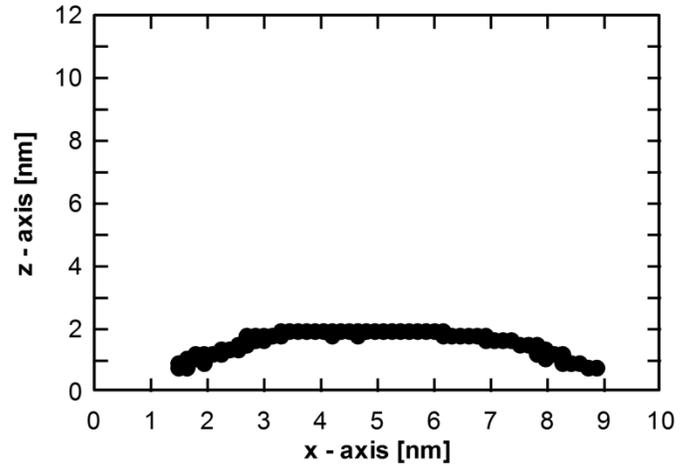

**Fig. 7. Detected droplet boundary**

If the centroid of the data was above the platinum surface then we should leave the z –coordinate of centroid as it is and angular search should be from $0 \leq \theta \leq 2\pi$. This will guarantee the contact angle estimation of droplet with contact angle greater than 90°. Also this is required if the droplet is floating in vacuum and not attached to the solid.

### Step 5: Solving for contact angle

With the detected LVB step 4, we can fit a circle using Landau method [20] and get the circle's equation which represents the droplet boundary. Landau method is an efficient method to fit a circle using non iterative geometric fit. This method relies on minimizing the error of fit between the set of points and the estimated arc. The main equations for this method are mentioned below. For a detailed version and nomenclature we encourage readers to see the original reference.

$$\bar{x} = \frac{c_1 b_2 - c_2 b_1}{a_1 b_2 - a_2 b_1} \tag{9}$$

$$\bar{y} = \frac{a_1 c_2 - a_2 c_1}{a_1 b_2 - a_2 b_1} \quad (10)$$

$$R^2 = \frac{1}{N}[\sum x^2 - 2\sum \bar{x} + N\bar{x}^2 + \sum y^2 - 2\sum \bar{y} + N\bar{y}^2] \quad (11)$$

Once we get the equation of circle we can solve it for the angle made by the tangent at the platinum-water interface. This step is graphically shown in Fig. 8.

During a fresh simulation it is advisable to do a test case and fine tune the parameters to make this algorithm well suited for the selected data. These parameters are including, but not limited to the weight functions of step 3, $z$ – cutoff of step 4.

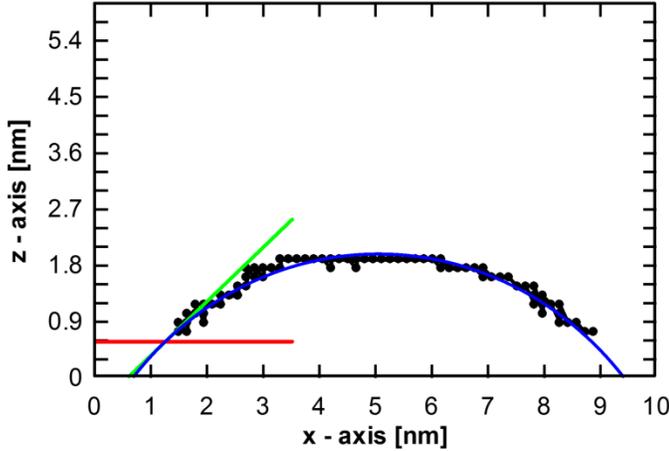

**Fig. 8. Final boundary of the droplet and fitted circle**

**CONTACT ANGLE ESTIMATION STUDIES**

**Case 1: Water-platinum interaction using LJ potential**

A cube shaped water droplet with 6 nm sides (7221 molecules) is kept on top of platinum wall interacting using LJ-potential. The sides are under periodic boundary condition and top is covered with another platinum surface to prevent the escape of molecules. The simulations are performed using Gromacs software. The platinum surfaces are modeled with FCC 111 structure and are square shaped with sides of 25 nm. The platinum surfaces are kept apart at a distance of 14 nm. There are three sub cases for this model based on the thermostat used. They are Berendsen, Nose-Hoover and Velocity re-scale thermostats. The simulation is performed from 0 ps to 1000 ps.

**Case 2: Water-platinum interaction using ZP-potential**

In this case a pre-equilibrated cubic shaped water droplet with sides of 4 nm (2133 molecules) is kept on top of FCC 111 platinum plate as shown in Fig. 9. The lateral boundaries are applied with periodic boundary condition (PBC) and vertical boundary is closed with platinum wall on bottom and mirror boundary on top. The platinum wall is of square shape with 20 nm sides and mirror boundary is kept at a distance of 12.5 nm from platinum. The water-platinum interaction is simulated using ZP-potential. A self-written and validated C++ code is used to perform the analysis. There are two sub cases of simulations based on this model. They are Berendsen and surface heating algorithm. More details of this simulation are described in the following paragraphs.

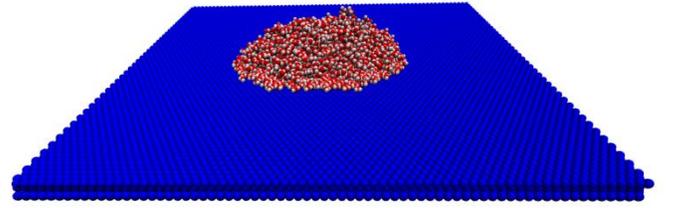

**Fig. 9. Simulation model for contact angle estimation**

The equations of motions are solved using velocity verlet scheme. A shifted scheme suggested by Stoddard and Ford [14] with 7σ was used instead of Ewald summation methods or P3M. The feature of this potential is that both potential and force goes smoothly to zero at the cut off radius.

$$U(r) = 4\varepsilon \left\{ \left[ \left(\frac{\sigma}{r}\right)^{12} - \left(\frac{\sigma}{r}\right)^{6} \right] \right.$$
$$\left. + \left[ 6\left(\frac{\sigma}{rc}\right)^{12} - 3\left(\frac{\sigma}{rc}\right)^{6} \right]\left(\frac{r}{rc}\right)^{2} - 7\left(\frac{\sigma}{rc}\right)^{12} + 4\left(\frac{\sigma}{rc}\right)^{6} \right\} \quad (12)$$

The water molecules are modeled using SPCE model introduced by Berendsen et al [14]. The intra molecular bonds are kept rigid throughout the simulation using the RATTLE algorithm [15].

Zhu-Philpott [16] model is used to model the interaction between water and platinum molecules. The advantage of such model is to obtain reasonable droplet shape formation during the simulation [4].

$$E_{Total} = E_{Conduction} + E_{Isotropic} + E_{Anisotropic} \quad (13)$$

$$E_{Conduction} = \sum_{real,image} \frac{q_{real} q_{image}}{2r} \quad (14)$$

$$E_{Isotropic} = -4\varepsilon_{w-Pt} \sum_{i=1}^{N_{Pt}} \frac{C_{w-Pt}\sigma_{w-Pt}^{10}}{r_{wi}^{10}} \quad (15)$$

$$E_{Anisotropic} = 4\varepsilon_{w-Pt} \sum_{i=1}^{N_{Pt}} \left[ \left( \frac{\sigma_{w-Pt}^2}{(\alpha\rho_{wi})^2 + z_{wi}^2} \right)^6 - \left( \frac{\sigma_{w-Pt}^2}{(\rho_{wi}/\alpha)^2 + z_{wi}^2} \right)^3 \right] \quad (16)$$

The parameters are taken from Maruyama's paper [6].

For simulations with thermostat based heating we have used Berendsen thermostat [1]. For those with wall heating we have used the modified Maroo-Chung model [3]. The details of this surface heating algorithm will be discussed in a separate paper which is yet to be published.

Since we are not modeling a contiguous array of droplets we can ignore the long range effects of the single water droplet. Simulation of droplet and films with SPME method will be studied extensively and will be published in another work. Hence we limit our studies with truncated potentials for the present work.

For the sub cases we have equilibrated the system at 300K for 200 ps. From 200 ps to 1000 ps the Berendsen thermostat or surface heating is applied for the two sub cases. For the last sub case we left the system uncoupled from thermostat from 200 ps to 1000 ps.

At the end of simulation the trajectory files are processed and contact angles are determined using the algorithm mentioned in the next section. The time evolution of contact angle is shown in Fig. 11 and Fig. 10.

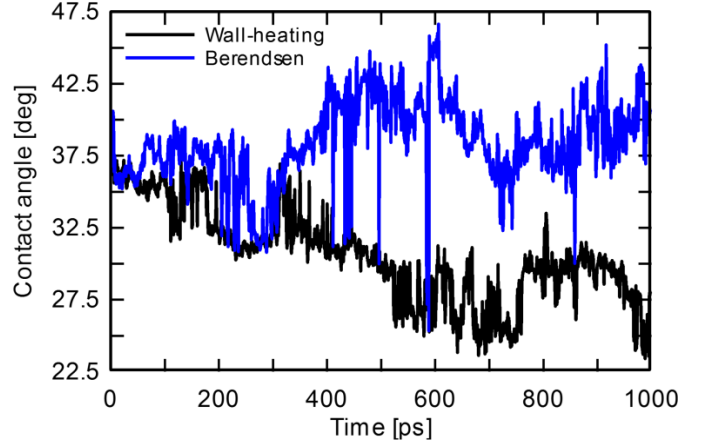

**Fig. 11. Contact angle evolution for case 2**

Since the main focus of this paper is not the effect and causes of the contact angle we have not included those discussions here. However we have made such a detailed study and currently that work is under review elsewhere. A working version of MATLAB code for this algorithm can be obtained from authors upon request.

**CONCLUSION**

A new algorithm for contact angle estimation from transient data of MD simulation was discussed. The efficiency and accuracy of this algorithm was demonstrated using different types of simulations. The MD simulations were done for water platinum interaction with ZP potential and LJ potential. Further, we used different thermostatting schemes for the droplet and show how it influences the contact angle. The results also showed the influence of interaction potential and also the thermostatting method used. The algorithm can be used to study the transient behavior of droplets on different surfaces.

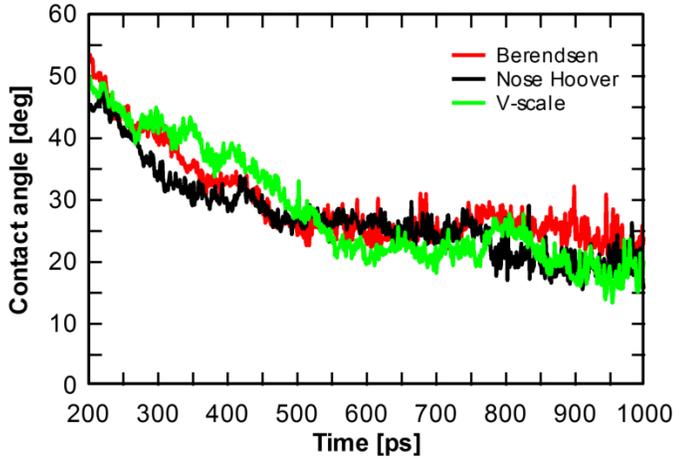

**Fig. 10. Contact angle evolution for case 1**